\documentclass[a4paper, oneside, twocolumn, notitlepage, 10pt]{extarticle_ecoc}
\usepackage{ecoc}

\usepackage{pgfplots}
\usepgfplotslibrary{groupplots}

\usepackage{color}
\usepackage{xcolor}
\usepackage{graphicx}
\usepackage{tikzscale}
\usepackage{tikz}
\usepackage{pgfplots}
\usetikzlibrary{plotmarks,matrix,chains,scopes,fit,calc,shapes,positioning,decorations,intersections,fit,backgrounds,patterns}
\usetikzlibrary{fadings,shapes.arrows,shadows}
\pgfplotsset{compat=newest} 

\pgfplotsset{plot coordinates/math parser=false}
\pgfplotsset{every axis plot/.append style={solid,line width=1.5pt,mark size=1.5pt,mark options={solid,fill=white}}}
\pgfplotsset{every axis legend/.append style={legend cell align=left,font=\footnotesize}}

\colorlet{42GBd16QAM_color}{blue!80!white}
\colorlet{42GBd64QAM_color}{red!80!black}
\colorlet{64GBd16QAM_color}{orange!90!black}
\colorlet{64GBd64QAM_color}{green!50!black}

\pgfplotsset{64GBd64QAM/.style={color=64GBd64QAM_color,solid}}
\pgfplotsset{64GBd16QAM/.style={color=64GBd16QAM_color,dotted}}
\pgfplotsset{42GBd64QAM/.style={color=42GBd64QAM_color,dashed}}
\pgfplotsset{42GBd16QAM/.style={color=42GBd16QAM_color,dash dot dot}}

\newlength\FigureWidth
\newlength\FigureHeight
\newlength\FullFigureWidth
\graphicspath{{figures/}}

\pgfplotsset{myLegend/.append style={legend style={font=\footnotesize,at={(0.5,0.98)},anchor=north,align=left,legend columns=3}}}

\usepackage{xparse}
\tikzfading[name=arrowfading, top color=transparent!0, bottom color=transparent!95]
\tikzset{arrowfill/.style={#1,general shadow={fill=black, shadow yshift=-0.8ex, path fading=arrowfading}}}
\tikzset{arrowstyle/.style n args={3}{draw=#2,arrowfill={#3}, single arrow,minimum height=#1, single arrow,
single arrow head extend=.3cm,}}

\NewDocumentCommand{\tikzfancyarrow}{O{2cm} O{FireBrick} O{top color=OrangeRed!20, bottom color=Red} m}{
\tikz[baseline=-0.5ex]\node [arrowstyle={#1}{#2}{#3}] {#4};
} 

\makeatletter
\tikzset{
    block filldraw/.style={
        draw, fill=yellow!20},
    block rect/.style={
        block filldraw, rectangle},
    block/.style={
        block rect, minimum height=0.8cm, minimum width=6em},
    from/.style args={#1 to #2}{
        above right={0cm of #1},
        /utils/exec=\pgfpointdiff
            {\tikz@scan@one@point\pgfutil@firstofone(#1)\relax}
            {\tikz@scan@one@point\pgfutil@firstofone(#2)\relax},
        minimum width/.expanded=\the\pgf@x,
        minimum height/.expanded=\the\pgf@y}}
\makeatother

\newcommand{\Ptx}{\ensuremath{P_{\text{tx}}}\xspace}

\newcommand{\SNR}{\ensuremath{\text{SNR}}}

\usepackage{amsmath,amssymb}

\begin{document}
\selectlanguage{english}    

\title{A QoT Estimation Method using EGN-assisted\\Machine Learning for Network Planning Applications}%
\author{
    Jasper M\"uller\textsuperscript{(1)},
    Sai Kireet Patri\textsuperscript{(1,2)},
    Tobias Fehenberger\textsuperscript{(1)},\\
    Carmen Mas-Machuca\textsuperscript{(2)},
    Helmut Griesser\textsuperscript{(1)},
    J\"org-Peter Elbers\textsuperscript{(1)}
}

\maketitle                  


\begin{strip}
 \begin{author_descr}

   \textsuperscript{(1)} ADVA, Fraunhoferstr. 9a, 82152 Martinsried/Munich, Germany,\textbf{}
   \textcolor{blue}{\uline{jmueller@adva.com}}

   \textsuperscript{(2)} Chair of Communication Networks, Technical University of Munich, Arcisstr. 21, Munich, Germany

 \end{author_descr}
\end{strip}

\setstretch{1.1}
\renewcommand\footnotemark{}
\renewcommand\footnoterule{}
\let\thefootnote\relax\footnotetext{978-1-6654-3868-1/21/\$31.00 \textcopyright 2021 IEEE}


\begin{strip}
  \begin{ecoc_abstract}
    An ML model based on precomputed per-channel SCI is proposed. Due to its superior accuracy over closed-form GN, an average SNR gain of 1.1 dB in an end-to-end link optimization and a 40\% reduction in required lightpaths to meet traffic requests in a network planning scenario are shown.
  \end{ecoc_abstract}
\end{strip}


\section{Introduction}
\vspace{-4pt}
With the advent of flexible grid bandwidth-variable transponders and low margin optical networking, network planning 
and optimization is becoming increasingly complex.
Simultaneously, driven by a rapidly growing demand in network throughput\cite{cisco}, the optimization of the current network infrastructure is of pivotal importance.
Modern routing, modulation and spectrum assignment algorithms require a path calculation engine (PCE), capable of fast and accurate quality of transmission (QoT) estimation.

The crucial part for QoT estimation is computing the nonlinear interference (NLI), for which a variety of models are available. These models, such as the split-step simulations and various Gaussian noise (GN) models, follow a trade-off between accuracy and complexity. 
In a C-band system using standard single mode fiber, the main NLI components are self-channel interference (SCI) and cross-channel interference (XCI). The vast majority of the computational complexity of accurate models like the enhanced GN (EGN) model\cite{egn} comes from the XCI computation.

Machine learning (ML) has become a popular tool\cite{Pointurier} for QoT estimation due to its ability to accurately predict nonlinear functions as well as its fast computation after the initial training. 
Most ML approaches in the literature fall into two classes. Classification methods have shown high accuracy in predicting the feasibility of a candidate lightpath in relation to a BER threshold \cite{Aladin, Rottondi, Morais}. Regression approaches, on the other hand, have been used for Q-factor prediction of multiple channels on a testbed link \cite{Gao}, in the context of modeling parameter uncertainty \cite{Pesic}, and recently for the estimation of a generalized signal-to-noise ratio (GSNR) distribution, assuming imperfect representation of physical parameters by the 
ML features\cite{Ibrahimi}. A neural-network-based NLI regression model has been demonstrated to accurately predict QoT in a live production network\cite{ofc}.

In this work, we propose a novel QoT estimation approach in which an ML regression model is using precomputed SCI values of each WDM channel as its input features in order to fast and accurately determine the total NLI for all channels. The capabilities of this approach using a gradient-boosting ML model are demonstrated in two proof-of-concept network planning scenarios. In a multi-step planning on a small network,  we achieve savings of over 40\% of the required lightpaths, and the spectral assignment optimization on an end-to-end link leads to an improvement of 1.1~dB in average SNR. 

\vspace{-8pt}
\section{Divide-and-conquer QoT Estimation} 
\vspace{-4pt}
SNR as figure of merit for QoT over a nonlinear fiber is defined as \cite{egn}
\begin{equation}
\setlength{\abovedisplayskip}{5pt}
\setlength{\belowdisplayskip}{5pt}
\SNR=\frac{\Ptx}{\sigma^2_\text{noise}}=\frac{\Ptx}{\sigma^2_\text{ASE} + \sigma^2_\text{SCI} + \sigma^2_\text{XCI}},    
\end{equation}
where multi-channel interference is neglected
. The computation of the total noise contribution can be split into three parts. The ASE noise can be calculated fast and straightforward, using well-known analytical formulas. The SCI computation is reasonably fast, even when using the full-form integral-based EGN model, because it requires integration over a small frequency band as SCI depends only on the channel under test (CUT) itself. In contrast, the XCI evaluation is computationally highly complex as it needs to consider all WDM channels and thus involves several THz of integration bandwidth.

\begin{figure}[t]
\centering
\includegraphics[width=\columnwidth]{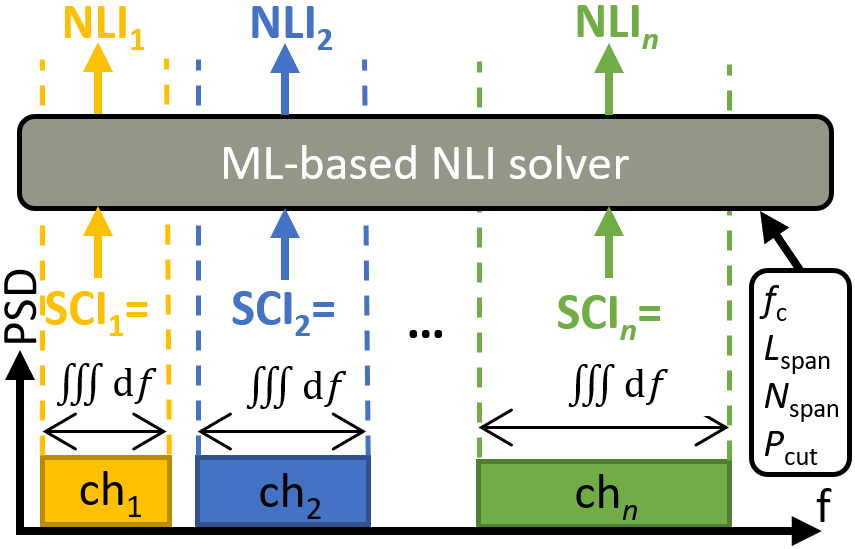}
\caption{Illustration of the proposed divide-and-conquer method for QoT estimation.}
\label{fig:NLI_Engine}
\end{figure}

The key idea of the proposed \emph{divide-and-conquer} approach for NLI computation is to compute only the SCI of each channel with the full-form EGN model ("divide") and store the SCI values for future use. An ML-based NLI solver uses the precomputed SCI values and some further inputs to determine the total NLI and thus the QoT of each lightpath ("conquer"). An illustration of the principle is shown in Fig.~\ref{fig:NLI_Engine}. The proposed scheme has to solve relatively simple integrals only once. Stored SCI values are reused for future NLI evaluation as in our model adding new lightpaths does not change the SCI of an existing channel. This is in contrast to conventional accurate NLI solvers that require taking into account the full spectrum to determine the NLI of each channel. The physical motivation behind our approach is that the SCI of a particular WDM channel depends on a set of parameters which largely define the XCI that this channel inflects onto its neighbors. Hence, instead of describing an interferer by its full spectrum (as we would do in the EGN model), the SCI of such an interferer serves as a proxy. The evaluation of XCI, and thus total NLI, based on SCI is done in an ML model that is described in the following.

\vspace{-8pt}
\section{ML-based NLI Solver}
\vspace{-4pt}
For computing NLI, a gradient-boosting ML model based on the XGBoost (XGB) \cite{xgb} library is used. In XGB, a decision tree ensemble is constructed one by one by fitting a new tree to the prediction error of the existing ones. XGB models are powerful ML models that are widely considered to be a preferable choice for tabular datasets. The advantage of the XGB model over neural networks frequently used for QoT estimation is that it is easier to optimize the model's hyperparameters and the feature selection. The output of a gradient-boosting model can be made explainable by tools such as TreeExplainer\cite{TreeExplainer}, quantifying the impact of a datapoint's input features on the model's output.
The XGB hyperparameters have been optimized on training data using cross-validation. The input features for the model were chosen from the available parameters using the XGB feature importance measuring the prediction power of the input features. Features with low prediction power were disregarded, optimizing the models performance on the validation set. The chosen input features are the SCI values of the CUT and its 10 closest WDM neighbors as well as their distance to the CUT. Additionally the CUT launch power, total number of channels, span length and the number of spans are chosen as inputs. The model is trained to output the NLI constant $\eta_\text{NLI}$ in dB, defined as the total NLI power normalized by the transmit power cubed.


The EGN model is used to generate a large and accurate data set for the training, validation and testing of the XGB ML model. The parameter space is shown in Table~\ref{tab:table1}. The C-Band is filled from 75\% up to 95\% with the channel parameters drawn uniformly. Frequency slots are chosen with a granularity of 12.5~GHz. The power spectral density (PSD) is assumed to be equalized over all channels, with a 100G QPSK channel at 35 GBd having 0~dBm launch power. In total over 230k data points were generated, representing more than 2200 different link configurations. The generated data set was split into training, validation and test set using an 70/10/20 split. Note that the data set has widely varying WDM grid layouts but only a limited number of link layouts. We found\cite{ofc} that links of heterogeneous span lengths and varying attenuation can be well represented by using combined parameters, such as the average of the cumulative sum of the span lengths, as the input into an ML model. 

\begin{table}[t]
   \centering
\caption{Parameter space for data generation} \label{tab:table1}
\begin{tabular}{|l|c|}
         \hline  $L_\text{span}$ [km]  & \{60, 80, 100, 120\}\\
         \hline  $N_\text{span}$  & 1 to 50, step=1\\
         \hline  Modulation  & QPSK, 16/32/64 QAM\\
         \hline  Data rate [GBit/s]  & 100 to 600, step=50\\
         \hline
\end{tabular}
\vspace{-\baselineskip}
\end{table}

\vspace{-8pt}
\section{Numerical Results on Simulation Data}
\begin{figure}[t]
\centering
\begin{tikzpicture}[font=\small]
\begin{axis}[
width=0.9\columnwidth,
height=0.6\columnwidth,
legend cell align={left},
legend style={
  fill opacity=0.8,
  draw opacity=1,
  text opacity=1,
  at={(0.97,0.03)},
  anchor=south east,
  draw=white!80!black
},
xlabel={SNR deviation [dB]},
ylabel={CDF},
xmin=0, xmax=0.8,
ymin=0.2, ymax=1,
xmajorgrids,
ymajorgrids,
xlabel shift = -5pt
]
\addplot [thick, red!60!black]
table {%
0.00630008352778333 0.0626339489970381
0.00734498870806988 0.0722662364244949
0.00825353478168456 0.0818985238519517
0.00917396345186461 0.0915308112794086
0.0101216566851292 0.101163098706865
0.0111880915354998 0.110795386134322
0.0121364858059048 0.120427673561779
0.0130805531608615 0.130059960989236
0.0140417212667305 0.139692248416693
0.015125532578713 0.14932453584415
0.0161985737966823 0.158956823271606
0.0172090060646255 0.168589110699063
0.0182629422123988 0.17822139812652
0.0193566272989063 0.187853685553977
0.0203321256238667 0.197485972981434
0.0213972125079991 0.207118260408891
0.0225227818166287 0.216750547836347
0.0234829496781739 0.226382835263804
0.0245216772568142 0.236015122691261
0.0256387072810256 0.245647410118718
0.0267744456212569 0.255279697546175
0.0278337193581741 0.264911984973632
0.0289834170817151 0.274544272401088
0.0301207165333413 0.284176559828545
0.0313770114544276 0.293808847256002
0.0325437090177587 0.303441134683459
0.033788462319972 0.313073422110916
0.0348317176322013 0.322705709538373
0.0360564302550088 0.332337996965829
0.0372551127998193 0.341970284393286
0.0385457704948369 0.351602571820743
0.0397585099286122 0.3612348592482
0.0410145711806891 0.370867146675657
0.0423585466845733 0.380499434103114
0.0436599307494792 0.39013172153057
0.0449602801399642 0.399764008958027
0.0462272647179853 0.409396296385484
0.0475764103093077 0.419028583812941
0.0488546888905645 0.428660871240398
0.0502384481784173 0.438293158667855
0.0516032025673745 0.447925446095311
0.0529508714190143 0.457557733522768
0.0544845958284537 0.467190020950225
0.0559292135228056 0.476822308377682
0.0574551128535532 0.486454595805139
0.0590532825979544 0.496086883232596
0.0607086965022816 0.505719170660052
0.062441157913744 0.515351458087509
0.0639898750389101 0.524983745514966
0.0656927755363412 0.534616032942423
0.0675202847672978 0.54424832036988
0.0693585457680577 0.553880607797337
0.0711249541342092 0.563512895224794
0.0731635359041984 0.57314518265225
0.0752824696125884 0.582777470079707
0.0771993301429754 0.592409757507164
0.0793425559659129 0.602042044934621
0.0814640554276949 0.611674332362078
0.0836280629908721 0.621306619789534
0.0857921480709036 0.630938907216991
0.0879521203788673 0.640571194644448
0.0903981915506149 0.650203482071905
0.0926952413589888 0.659835769499362
0.0953715698558817 0.669468056926819
0.0977629342764423 0.679100344354276
0.100560648823347 0.688732631781732
0.10332201136443 0.698364919209189
0.106045022023725 0.707997206636646
0.108954551876483 0.717629494064103
0.112044734865628 0.72726178149156
0.115155068579647 0.736894068919017
0.118382489561668 0.746526356346473
0.122086006113586 0.75615864377393
0.125736882383411 0.765790931201387
0.12926594337732 0.775423218628844
0.133176070145288 0.785055506056301
0.13762658330462 0.794687793483758
0.141436229886249 0.804320080911214
0.145691173597271 0.813952368338671
0.150240477995631 0.823584655766128
0.155060169463262 0.833216943193585
0.159963660136823 0.842849230621042
0.165127864580462 0.852481518048499
0.170924904340982 0.862113805475955
0.176488924118113 0.871746092903412
0.182862895033951 0.881378380330869
0.190104407113487 0.891010667758326
0.197845082853497 0.900642955185783
0.206819248553453 0.91027524261324
0.216426235836941 0.919907530040696
0.226788017947838 0.929539817468153
0.238048550502926 0.93917210489561
0.252343079557726 0.948804392323067
0.26864380305113 0.958436679750524
0.290243338597719 0.968068967177981
0.319732648802288 0.977701254605437
0.36286884753153 0.987333542032894
0.407579439262381 0.993112914489368
0.417884049121092 0.994076143232114
0.430184438132704 0.99503937197486
0.45195980598252 0.996002600717605
0.476646198744398 0.996965829460351
0.511409978232869 0.997929058203097
0.557510833159197 0.998892286945842
0.820859569873088 0.999855515688588
};
\addlegendentry{ML}

\addplot [thick, dashed, green!40!black]
table {%
1.90515159204097e-07 4.81609732369472e-06
0.00284987710331386 0.00963701074471313
0.00552103067173171 0.0192692053921026
0.00825569338272913 0.028901400039492
0.0110369015674774 0.0385335946868814
0.0138506464022186 0.0481657893342709
0.0166216128109404 0.0577979839816603
0.0192455605514468 0.0674301786290497
0.0219619254455239 0.0770623732764392
0.0245793973411566 0.0866945679238286
0.0271777787316285 0.096326762571218
0.0297659904509437 0.105958957218607
0.0322958063104393 0.115591151865997
0.0348384766544285 0.125223346513386
0.0373728650873684 0.134855541160776
0.0398284739418528 0.144487735808165
0.0421997212729543 0.154119930455555
0.044583581349471 0.163752125102944
0.0468735995968252 0.173384319750334
0.0491403274394671 0.183016514397723
0.051419822402611 0.192648709045112
0.0535954975901216 0.202280903692502
0.0557259702615447 0.211913098339891
0.057893557119769 0.221545292987281
0.0600463283802632 0.23117748763467
0.0622127023601777 0.24080968228206
0.064326587753416 0.250441876929449
0.0663856138558634 0.260074071576838
0.0684523666793808 0.269706266224228
0.0705249783198942 0.279338460871617
0.0725660160169195 0.288970655519007
0.0746343054864482 0.298602850166396
0.0766363640277739 0.308235044813786
0.0786824507261041 0.317867239461175
0.0806599693075274 0.327499434108564
0.0827444162682252 0.337131628755954
0.0848618113655295 0.346763823403343
0.0868657998769411 0.356396018050733
0.0888768602401964 0.366028212698122
0.0909444760415834 0.375660407345512
0.0930585195489684 0.385292601992901
0.095127191014079 0.394924796640291
0.0971716759621035 0.40455699128768
0.0993710337402511 0.414189185935069
0.101560630269866 0.423821380582459
0.103768089399995 0.433453575229848
0.105933528786079 0.443085769877238
0.108144918088087 0.452717964524627
0.11040502101384 0.462350159172017
0.112702624345975 0.471982353819406
0.115062251324161 0.481614548466795
0.117338357920064 0.491246743114185
0.11967783441464 0.500878937761574
0.122084489052932 0.510511132408964
0.124569950533932 0.520143327056353
0.127095448941098 0.529775521703743
0.129741472207174 0.539407716351132
0.132368249287984 0.549039910998521
0.135046510254082 0.558672105645911
0.137828633765489 0.5683043002933
0.140672768521346 0.57793649494069
0.143694434899665 0.587568689588079
0.146661589904696 0.597200884235469
0.149690278490256 0.606833078882858
0.152810788501082 0.616465273530247
0.156064220437178 0.626097468177637
0.159445587898716 0.635729662825026
0.162877562176462 0.645361857472416
0.16621465453027 0.654994052119805
0.169865109408754 0.664626246767195
0.173502314853136 0.674258441414584
0.177167528669079 0.683890636061974
0.180923552526078 0.693522830709363
0.184795773945957 0.703155025356752
0.188870003712463 0.712787220004142
0.193016231275168 0.722419414651531
0.197166928041323 0.732051609298921
0.201527077235166 0.74168380394631
0.206103578182566 0.7513159985937
0.210812031765915 0.760948193241089
0.215739748292176 0.770580387888478
0.220669338707079 0.780212582535868
0.225807420720226 0.789844777183257
0.231124576623927 0.799476971830647
0.236688082327301 0.809109166478036
0.242452438657754 0.818741361125426
0.248485493389126 0.828373555772815
0.2548605805738 0.838005750420205
0.261739306978017 0.847637945067594
0.268882842099796 0.857270139714983
0.276327531490555 0.866902334362373
0.284403096986992 0.876534529009762
0.292855857639008 0.886166723657152
0.302552095505945 0.895798918304541
0.312645067196542 0.905431112951931
0.323574826067516 0.91506330759932
0.335855638968017 0.924695502246709
0.349519911127073 0.934327696894099
0.364807634676978 0.943959891541488
0.383142349938344 0.953592086188878
0.406411327753961 0.963224280836267
0.437381121140426 0.972856475483657
0.483435613882932 0.982488670131046
0.577962532792199 0.992120864778435
0.595304152658258 0.993084084243174
0.616550189918662 0.994047303707913
0.640480348656144 0.995010523172652
0.670254387339522 0.995973742637391
0.706939806442172 0.99693696210213
0.756008398763463 0.997900181566869
0.821513520362643 0.998863401031608
0.960877646697085 0.999826620496347
};
\addlegendentry{GN}

\draw[<-, very thick, red!60!black] (0.095,0.7) -- (0.08,0.78) node[above,align=left] {ML\\mean};
\draw[<-, very thick, green!40!black] (0.15,0.6) -- (0.25,0.55) node[below] {GN mean};

\draw[<-, very thick, red!60!black] (0.45,0.99) -- (0.42,0.9) node[below] {ML 99th \%};
\draw[<-, very thick, green!40!black] (0.55,0.99) -- (0.6,0.8) node[below] {GN 99th \%};
\end{axis}

\end{tikzpicture}

\vspace{-.3\baselineskip}
\caption{CDF of the SNR deviation between ML model (red) and GN (green, dashed), in relation to EGN.}
\label{fig:cdf}
\end{figure}
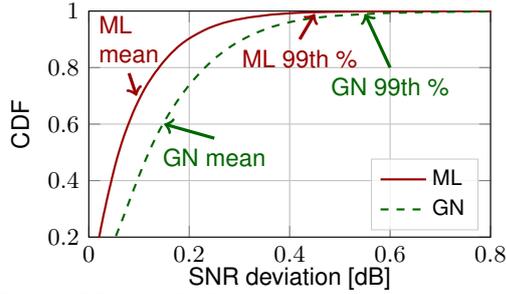

\vspace{-4pt}
 In Fig.~\ref{fig:cdf}, the cumulative distribution function (CDF) of the SNR error in dB of the ML model is compared to that of the closed-form EGN approximate model \cite{Zefreh}, henceforth referred to as GN. A mean error of less than 0.1~dB SNR is obtained for the ML-based QoT estimator and 0.15~dB for the GN. On 99\% of the test data set, the ML models error lies below 0.45 dB SNR, whereas the 99th percentile for the GN is 0.55 dB. A clear advantage of the ML model is its computation time, which is just 17~\textmu s (i7-7500U CPU with 12GB RAM), and thus about 300~times faster than the 5 ms needed for the GN. In comparison, the EGN model takes on average 100 seconds. Although the millisecond time scale and the accuracy of the GN might be sufficient for a variety of applications, complex network planning tasks could require the performance and in particular the speed improvement offered by the ML model. In conclusion, the proposed ML-based QoT estimator is orders of magnitude faster than the EGN model with a prediction accuracy that is comparable to previous QoT estimators\cite{ofc}. In the following, it is demonstrated in two proof-of-concept application scenarios.
 


\begin{figure}[ht]
\centering
  \input{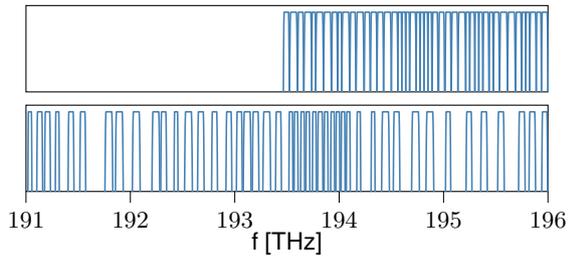}

\vspace{-1.2\baselineskip}
\caption{Spectra of first fit (top) and ML optimization (bottom), resulting in 1.1~dB average SNR difference.}
\label{fig:spec_opt}
\end{figure}

\vspace{-8pt}
\section{Application 1: Spectral Optimization}
\vspace{-4pt}
The first application is the optimization of the spectral assignment on a 960~km link. The 5~THz C-band was split into 400 slots of 12.5~GHz each. The spectrum was assumed to be filled to 50\%. As a baseline, channels of different configurations are placed in a first-fit manner, as shown in Fig.~\ref{fig:spec_opt}.


For the same order of demands to be placed, the optimization algorithm using the ML model chooses the slot that minimizes the total NLI of all previous channels by sweeping over all possible placements. 
The optimized spectrum is shown at the bottom of Fig.~\ref{fig:spec_opt}. An SNR improvement was achieved for all channels, and the average gain of 1.1~dB corresponds to a considerable increase in system margin.
This brute-force optimization algorithm required a total of 
more than 120k 
NLI computations, which took 8 minutes. This demonstrates the need for fast and accurate QoT estimators in order to optimize the current network infrastructure.

\vspace{-8pt}
\section{Application 2: PCE for Network Planning}
\vspace{-3pt}
In a multi-period scenario, a large number of QoT calculations is required. Therefore, full-form integral models like the EGN model cannot be applied. 
Open source PCEs like GNPy \cite{gnpy} use GNs in order to provide quick QoT estimates. However, these calculations suffer from estimation inaccuracies accumulated during multi-period planning. To mitigate this, we integrated our proposed ML-based QoT estimator into a multi-period network planner\cite{patri2020planning} and conducted a five-year network planning study for the small NORDUnet network\cite{nordunet}, consisting of 5 nodes, 4 links, and 10 demand pairs. We assume that each demand pair generates traffic requests~(TRs) every year according to a recently proposed traffic model\cite{patri2020planning}. The RMSA algorithm then places lightpaths in order to meet the demands. This exercise is repeated using ML and GN as PCE.

\begin{figure}[t]
\centering
  \begin{tikzpicture}[font=\small]

\definecolor{color0}{rgb}{0.549019607843137,0,0.0588235294117647}
\definecolor{color1}{rgb}{0.0823529411764706,0.690196078431373,0.101960784313725}

\begin{groupplot}[group style={group size=2 by 1, horizontal sep=20pt}]
\nextgroupplot[
legend cell align={left},
legend style={
  fill opacity=0.8,
  draw opacity=1,
  text opacity=1,
  at={(0.03,0.97)},
  anchor=north west,
  draw=white!80!black
},
tick align=outside,
tick pos=left,
x grid style={white!69.0196078431373!black},
xlabel={Planning Years},
xmajorgrids,
xmin=-0.2, xmax=4.2,
xtick style={color=black},
xtick={0,1,2,3,4},
xticklabels={1,2,3,4,5},
y grid style={white!69.0196078431373!black},
ylabel={Lightpath Count},
ymajorgrids,
ymin=19, ymax=151,
ytick style={color=black},
width=0.45*\columnwidth,
height=0.5*\columnwidth,
ylabel shift = -6pt,
]
\addplot [thick, red!60!black]
table {%
0 25
1 30
2 52
3 80
4 103
};
\addplot [thick, green!40!black, dashed]
table {%
0 25
1 30
2 60
3 108
4 145
};
\nextgroupplot[
legend style={at={(0.995,0.995)}, anchor=north east},
tick align=outside,
tick pos=left,
x grid style={white!69.0196078431373!black},
xlabel={Lightpath datarate [Gbps]},
xmin=70, xmax=430,
xtick style={color=black},
xtick={100,150,200,250,300,350,400},
xticklabels={100,,200,,300,,400},
y grid style={white!69.0196078431373!black},
ymin=0, ymax=80,
ytick style={color=black},
width=0.7*\columnwidth,
height=0.5*\columnwidth,
bar width=0.17cm,
ybar,
legend image code/.code={%
      \draw[#1,draw=none] (0cm,-0.1cm) rectangle (0.6cm,0.1cm);
    }   
]



\addplot[draw=none, fill=red!60!black, solid, bar shift=-.09cm]
table[row sep=crcr] {%
100 2 \\ 
150 6 \\ 
200 47 \\ 
250 3 \\ 
300 35 \\ 
350 1 \\ 
400 9 \\ 
};
\addlegendentry{ML};

\addplot[draw=none, fill=green!40!black, postaction={
        pattern=north east lines, pattern color=green!20!black
    }, bar shift=+.09cm]
table[row sep=crcr] {%
100 55 \\ 
150 3 \\ 
200 75 \\ 
250 0 \\ 
300 2 \\ 
350 1 \\ 
400 9 \\ 
};
\addlegendentry{GN};

\end{groupplot}

\end{tikzpicture}

\vspace{-0.9\baselineskip}
\caption{Number of lightpaths used in multi-period planning scenario per year (left) and per configuration (right).}
\label{fig:networkplanning}
\end{figure}
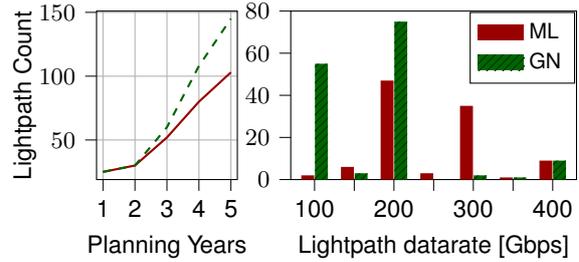

As shown in Fig.~\ref{fig:networkplanning}, the ML model suggests that a higher datarate can be used for the lightpaths in the network than when using the GN estimator. Since the planning algorithm tries to achieve the TR with minimal number of lightpaths, both QoT estimators are able to satisfy all TRs every year. However, ML is able to achieve the TRs by placing 40\% fewer lightpaths in the five year planning period, as shown in Fig.~\ref{fig:networkplanning}. This leads to an overall reduction in spectral usage, and can be helpful in delaying the need to light additional dark fibers.

\vspace{-8pt}
\section{Conclusions}
\vspace{-4pt}
We have introduced an ML-based QoT estimator, taking in SCI values computed by EGN. The ML model achieves high accuracy on simulation data (\textless 0.1dB mean SNR error), outperforming GN models. We demonstrated that using the proposed ML based QoT estimator, network planners can optimize spectral assignments to increase margins considerably (1.1~dB) and reduce the number of required lightpaths in a multi-period planning scenario by 40\%. Using an ML-based QoT estimator, the conservative calculations from closed-form GNs can be avoided.

\clearpage

\printbibliography

\end{document}